\documentclass[twocolumn,twoside,slac_two]{revtex4}
\usepackage{graphicx}
\usepackage{fancyhdr}
\begin{document}

\title{A Search for Candidate TeV Emitters in the High-latitude {\it{Fermi}} Unassociated Sources}

\author{P. Fortin}
\affiliation{Laboratoire Leprince Ringuet / \'Ecole Polytechnique / CNRS / IN2P3, Palaiseau, France}
\author{D. Horan}
\affiliation{Laboratoire Leprince Ringuet / \'Ecole Polytechnique / CNRS / IN2P3, Palaiseau, France}
\author{E. Ferrara}
\affiliation{NASA Goddard Space Flight Center, Greenbelt, MD, USA}
\author{on behalf of the \it{Fermi}-LAT Collaboration}

\begin{abstract}
We report the results of an analysis to identify candidates for
very-high-energy (VHE; E $>$ 100 GeV) emission from the high-latitude
($|b| > 10$) unassociated sources in the year-1 catalog under
development by the {\it{Fermi}} Large Area Telescope (LAT) team. These
are sources with no known counterparts at other wavelengths. Since VHE
instruments are pointed instruments with small fields of view and low
duty cycles, their observing programs need to be planned carefully to
identify the most promising targets for observation. The scientific
potential of combined {\it{Fermi}} and VHE observations has already
been demonstrated with a number of joint VHE-{\it{Fermi}} papers. The
goal of this work is to select the most promising unassociated
{\it{Fermi}} sources for joint observations with {\it{Fermi}} and the
VHE instruments.
\end{abstract}

\maketitle

\thispagestyle{fancy}

\section{Introduction}

The goal of this work is to find candidate very-high-energy (VHE)
emitters from the sources in the {\it{Fermi}} year-1
catalogue. Many of the sources that have already been detected
at VHE are in the catalog so it is already clear that {\it{Fermi}} can
provide further targets of interest for VHE observatories.

\section{Metholdology}

\subsection{Preliminary Scan}

To begin, a preliminary scan of the catalog was performed to select
unassociated sources above galactic latitudes of 10$^{\circ}$ that had
potential to be VHE emitters based on their flux, spectra and
redshift. Unassociated sources that met the following criteria were
selected:

\begin{itemize}
\item $|$ Galactic Latitude $|$ $>$ $10^{\circ}$
\item Flux (E\,$>$\,100\,MeV) $>$ 2\,$\times$\,10$^{-9}$\,cm$^{-2}$\,s$^{-1}$
\item Photon Index $< 2.0$
\item Number of predicted photons $>20$
\end{itemize}

\begin{figure}
\includegraphics[width=65mm]{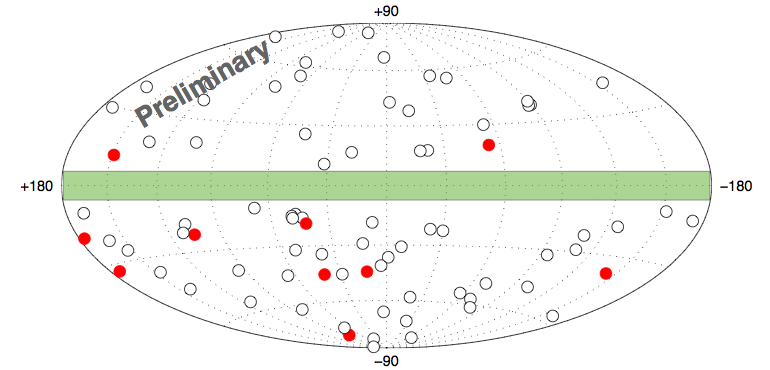}
\caption{Spatial distribution of the 80 candidates selected for this study in galactic coordinates. The solid red circles correspond to the top 10 candidates shown in Table \ref{TAB:list}. The green-shaded region corresponds to the $\pm 10^{\circ}$ exclusion region centered on the galactic plane.} \label{FIG:skymap}
\end{figure}

This resulted in a list of 80 candidates corresponding to $\sim$11\%
of the total number of unassociated sources in the year-1 catalog. A skymap of
these sources in galactic coordinates is shown in
Figure~\ref{FIG:skymap}.

\subsection{Detailed Analysis}

These 80 sources were then re-analyzed to search for curvature in
their spectra; the sources in the catalog were all fit with power-law
spectra, which gives, to first order, a good indication of their
properties across the {\it{Fermi}} energy band but does not always
describe accurately their spectra above 1\,GeV. We included only those
data at energies above 1\,GeV from these sources to see if the
resulting flux agrees with the power-law fit in the catalog. For each
source, events were selected from a region of 10$^\circ$ radius
centered on the catalog coordinates for that source. These data were
analyzed using the standard {\it{Fermi}} analysis software
(\texttt{ScienceTools v9r15p3; IRF P6\_V3\_DIFFUSE}) available from
the HEASARC. All of the {\it{Fermi}} sources in the field of view were
modeled and the background emission was modeled using a galactic
diffuse emission model and an isotropic component \cite{1}. Events
were analyzed using an unbinned maximum likelihood method \cite{2},
\cite{3}.

The sources were modeled using a power law covering two overlapping
energy ranges: 100\,MeV\,-\,300\,GeV and 1\,GeV\,-\,300\,GeV. In
addition, a log parabola covering the full energy range
(100\,MeV\,-\,300\,GeV) was used to search for curvature. The
differential flux, $F(E)$ as a function of energy, $E$, for the
power-law and the log-parabola spectral functions are shown in
equations~\ref{EQ:powerlaw} and \ref{EQ:logparabola}, respectively.

\begin{equation}\label{EQ:powerlaw}
  F(E)=\frac{dN}{dE}=F_0\left(\frac{E}{E_0}\right)^{-\Gamma}
\end{equation}

\begin{equation}\label{EQ:logparabola}
  F(E)=\frac{dN}{dE}=F_0\left(\frac{E}{E_b}\right)^{-({\alpha}+{\beta}ln\left(\frac{E}{E_b}\right))}
\end{equation}

\begin{figure}
\includegraphics[width=65mm]{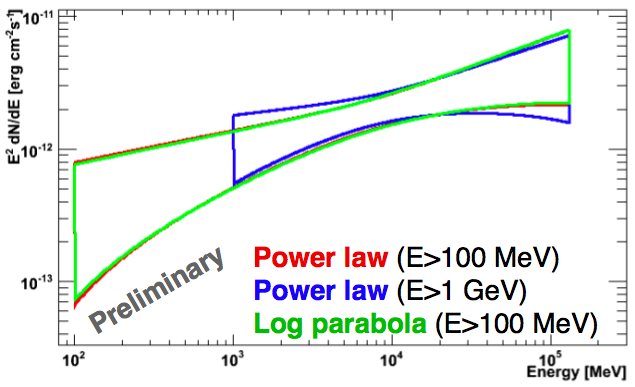}
\caption{SED of good TeV candidate showing no evidence for curvature in the \textit{Fermi} energy band.} \label{FIG:sedgood}
\end{figure}

\begin{figure}
\includegraphics[width=65mm]{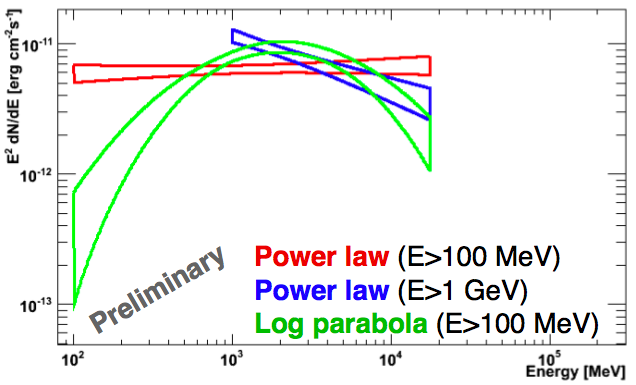}
\caption{SED of bad TeV candidate showing clear evidence for curvature.} \label{FIG:sedbad}
\end{figure}

$F_0$ is the differential flux at the decorrelation energy, $E_0$;
$\Gamma$ is the photon index; $E_b$ is the break energy; $\alpha$ and
$\beta$ are the parameters of the log-parabola fit. The likelihood
values of the power-law and log-parabola fits over the full energy
range were compared and a $\chi^2$ statistical test was used to
calculate the probability of curvature in the photon spectrum. The
covariance matrices were used to calculate the 1-sigma confidence
intervals of the spectral energy distribution (SED) for each spectral
model. The energy range of the SED starts at 100\,MeV and terminates
at the energy of the most energetic photon associated with each
source. Figures~\ref{FIG:sedgood} and \ref{FIG:sedbad} show SED
examples for sources with and without evidence for curvature.

\subsection{Predicting the TeV Flux}

For sources where there was no evidence for curvature, the flux above
100\,MeV and the spectral index obtained from the power-law fit were
used to estimate the flux in the VHE band (200\,GeV\,-\,1\,TeV). The
{\it{Fermi}} spectrum was extrapolated to higher energies assuming no
break in the spectrum and the flux was absorbed for the extragalactic
background light (EBL) with the model of Franceschini et
al. \cite{4}. A redshift value of $z$\,=\,$0.2$ was assumed for all
sources.

For those sources where a log-parabola was a better fit, the flux and
spectral index obtained using the power-law fit above 1\,GeV were used
for the extrapolation to the VHE band. Given that curvature was
detected in the {\it{Fermi}} energy band, it is likely that the
predicted TeV flux is overly optimistic. But, given the lack of
knowledge about the redshift of these sources, we consider the
brightest ones as good ``filler" targets in under-populated RA
bands. Table~\ref{TAB:list} lists the 10 most promising unassociated
{\it{Fermi}} sources for the VHE band using our methodology.

%\noindent{\textsc{\bf{References}}}

%\noindent(1) http:\/\/fermi.gsfc.nasa.gov/ssc/data/access/lat/BackgroundModels.html\\
%\noindent(2) Cash, W. 1979, ApJ, 228, 939\\
%\noindent(3) Mattox, J.R., et al. 1996, ApJ, 461, 396\\
%\noindent(4) Franceschini, A., Rodighiero, G., \& Vaccari, M. 2008, A\&A, 487, 837\\

\begin{table*}[htdp]
\begin{center}
{\footnotesize
\begin{tabular}{|l|c|c|c|c|c|c|c|c|c|}
\hline
Source & RA         & Dec         & Fermi Flux    & Fermi Index     & E$_{max}$  & \multicolumn{2}{c|}{TeV Flux}      \\
Name   & (J2000)    & (J2000)     & [$E>100$ MeV]        &                 & [GeV]      & $10^{-12} cm^{-2} s^{-1}$ & \%Crab \\
& & & [E-09] & & & & \\
\hline

TeV 01 & 00 22 22.1 & -18 48 42.9 & $4.7 \pm 1.4$ & $1.65 \pm 0.11$ & 96         & 10.0                      & 4 \\
TeV 02 & 03 38 59.5 & +13 12 39.7 & $4.1 \pm 2.2$ & $1.64 \pm 0.18$ & 133        & 8.91                      & 4 \\
TeV 03 & 21 18 18.4 & -32 38 27.5 & $3.3 \pm 1.7$ & $1.61 \pm 0.18$ & 35         & 8.64                      & 3 \\
TeV 04 & 05 05 52.4 & +61 22 32.7 & $4.1 \pm 2.4$ & $1.65 \pm 0.18$ & 131        & 8.39                      & 3 \\
TeV 05 & 13 07 40.4 & -43 00 19.9 & $13  \pm 4.1$ & $1.84 \pm 0.11$ & 31         & 7.77                      & 3 \\
TeV 06 & 23 29 13.4 & +37 55 29.9 & $6.5 \pm 2.5$ & $1.74 \pm 0.13$ & 71         & 7.51                      & 3 \\
TeV 07 & 04 39 53.6 & -18 58 02.3 & $3.3 \pm 1.5$ & $1.65 \pm 0.17$ & 49         & 6.65                      & 3 \\
TeV 08 & 21 46 40.8 & -13 45 30.3 & $7.9 \pm 3.1$ & $1.79 \pm 0.15$ & 52         & 6.60                      & 3 \\
TeV 09 & 20 14 26.0 & -00 45 53.2 & $2.5 \pm 2.2$ & $1.65 \pm 0.28$ & 114        & 5.39                      & 2 \\
TeV 10 & 04 27 21.6 & +20 26 06.0 & $2.2 \pm 2.2$ & $1.64 \pm 0.29$ & 71         & 4.89                      & 2 \\

\hline
\end{tabular} }
\end{center}
\caption{Predicted fluxes in the VHE energy band (200\,GeV\,-\,1\,TeV) for a the top 10 candidates for VHE emission from the {\it{Fermi}} unassociated sources in the year-1 catalog.}
\label{TAB:list}
\end{table*}%	

\subsection{Acknowledgments} \label{Ack}

The {\it{Fermi}} LAT Collaboration acknowledges the generous support
of a number of agencies and institutes that have supported the
{\it{Fermi}} LAT Collaboration. These include the National Aeronautics
and Space Administration and the Department of Energy in the United
States, the Commissariat \`a l'Energie Atomique and the Centre
National de la Recherche Scientifique / Institut National de Physique
Nucl\'eaire et de Physique des Particules in France, the Agenzia
Spaziale Italiana and the Istituto Nazionale di Fisica Nucleare in
Italy, the Ministry of Education, Culture, Sports, Science and
Technology (MEXT), High Energy Accelerator Research Organization (KEK)
and Japan Aerospace Exploration Agency (JAXA) in Japan, and the
K.~A. Wallenberg Foundation, the Swedish Research Council and the
Swedish National Space Board in Sweden.

Additional support for science analysis during the operations phase
from the following agencies is also gratefully acknowledged: the
Istituto Nazionale di Astrofisica in Italy and the K.~A. Wallenberg
Foundation in Sweden.

This research has made use of the NASA/IPAC Extragalactic Database
(NED) which is operated by the Jet Propulsion Laboratory, California
Institute of Technology, under contract with the National Aeronautics
and Space Administration. This research has made use of the SIMBAD
database, operated at CDS, Strasbourg, France.

The authors wish to thank JACoW for their guidance in preparing
this template. Work supported by Department of Energy contract DE-AC03-76SF00515.

\bigskip % extra skip inserted

\end{document}